\documentclass[12pt]{article}
\usepackage{times}
\usepackage{geometry}
\usepackage[utf8]{inputenc}
\usepackage{enumitem,amssymb}
\usepackage{ragged2e}
\newlist{thematic}{itemize}{8}
\setlist[thematic]{label=$\square$}
\usepackage{pifont}
\usepackage{wrapfig}
\usepackage[pdftex]{graphicx}

\begin{document}
\raggedright
\huge
Astro2020 Science White Paper \linebreak

Triggered High-Priority Observations of Dynamic Solar System Phenomena \linebreak
\normalsize

\noindent \textbf{Thematic Areas:} \hspace*{60pt} X Planetary Systems \hspace*{10pt} $\square$ Star and Planet Formation \hspace*{20pt}\linebreak
$\square$ Formation and Evolution of Compact Objects \hspace*{31pt} $\square$ Cosmology and Fundamental Physics \linebreak
  $\square$  Stars and Stellar Evolution \hspace*{1pt} $\square$ Resolved Stellar Populations and their Environments \hspace*{40pt} \linebreak
  $\square$    Galaxy Evolution   \hspace*{45pt} $\square$             Multi-Messenger Astronomy and Astrophysics \hspace*{65pt} \linebreak
  
\textbf{Principal Author:}

Name:	Nancy Chanover
 \linebreak						
Institution:  New Mexico State University
 \linebreak
Email: nchanove$@$nmsu.edu
 \linebreak
Phone:  575-646-2567
 \linebreak
 
\textbf{Co-authors:}\\
Michael H. Wong, UC Berkeley\\
Thomas Greathouse, SwRI\\
David Trilling, Northern Arizona University\\
Al Conrad, LBTO\\
Imke de Pater, UC Berkeley\\
Eric Gaidos, University of Hawai`i\\
Richard Cartwright, SETI Institute\\
Michael Lucas, U Tennessee Knoxville\\
Karen Meech, University of Hawai`i\\
Glenn Orton, Jet Propulsion Laboratory\\
Noemi Pinilla-Alonso, Florida Space Institute (UCF) \& Arecibo Observatory\\
Kunio Sayanagi, Hampton University\\
Megan E. Schwamb, Gemini Observatory\\
Matthew Tiscareno, SETI Institute\\
Christian Veillet, LBTO\\
Bryan Holler, STScI\\
Katherine de Kleer, Caltech\\
Heidi Hammel, AURA\\
Amanda Hendrix, Planetary Science Institute\\
Angel Otarola, TMT International Observatory\\
Conor Nixon, NASA Goddard Space Flight Center\\
Susan Benecchi, Planetary Science Institute\\
Amy Simon, NASA Goddard Space Flight Center\\
Kathleen Mandt, Johns Hopkins University Applied Physics Laboratory\\
Anne Verbiscer, University of Virginia\\
Rohini Giles, SwRI\\
Kurt Retherford, SwRI\\
Patrick Fry, University of Wisconsin - Madison\\
James F. Bell, Arizona State University\\
Stefanie Milam, NASA Goddard Space Flight Center\\
Andy Rivkin, Johns Hopkins University Applied Physics Laboratory\\
Statia Luszcz-Cook, Columbia University \& AMNH
\vspace{0.1in}


\textbf{Abstract:}

Unexpected dynamic phenomena have surprised solar system observers in the past and have led to important discoveries about solar system workings. Observations at the initial stages of these events provide crucial information on the physical processes at work. We advocate for long-term/permanent programs on ground-based and space-based telescopes of all sizes -- including Extremely Large Telescopes (ELTs) -- to conduct observations of high-priority dynamic phenomena, based on a predefined set of triggering conditions. These programs will ensure that the best initial dataset of the triggering event are taken; separate additional observing programs will be required to study the temporal evolution of these phenomena. While not a comprehensive list, the following are notional examples of phenomena that are rare, that cannot be anticipated, and that provide high-impact advances to our understandings of planetary processes. Examples include:
\begin{itemize}
  \item new cryovolcanic eruptions or plumes on ocean worlds
  \item impacts on Jupiter, Saturn, Uranus, or Neptune
  \item extreme eruptions on Io
  \item convective superstorms on Saturn, Uranus, or Neptune
  \item collisions within the asteroid belt or other small-body populations
  \item discovery of an interstellar object passing through our solar system ({\it e.g.} `Oumuamua)
  \item responses of planetary atmospheres to major solar flares or coronal mass ejections
\end{itemize}
\pagebreak

Observing programs triggered by new and unforeseen solar system discoveries and phenomena can exploit the unique capabilities of ground-based telescopes of all sizes for making time-critical followup observations. In some cases these will require the advanced capabilities of ELTs such as the Thirty Meter Telescope (TMT) and the Giant Magellan Telescope (GMT), as well as advanced large-aperture space telescopes. {\it The overarching goal of these observations is to use these dynamic phenomena to understand key processes that have shaped the formation and evolution of our solar system.} This goal is directly aligned with the range of big questions outlined in the most recent astrophysics and planetary science decadal surveys and the NASA science plan.
We describe several such opportunistic observations made over the past decade using current state-of-the-art facilities.  These are meant to serve as {\it examples} of triggered solar system observations, with the recognition that there are likely to be new serendipitous discoveries equally appropriate for triggered follow-up observations.\\



\begin{figure}[h]
    \centering
    \includegraphics[width=6.5in]{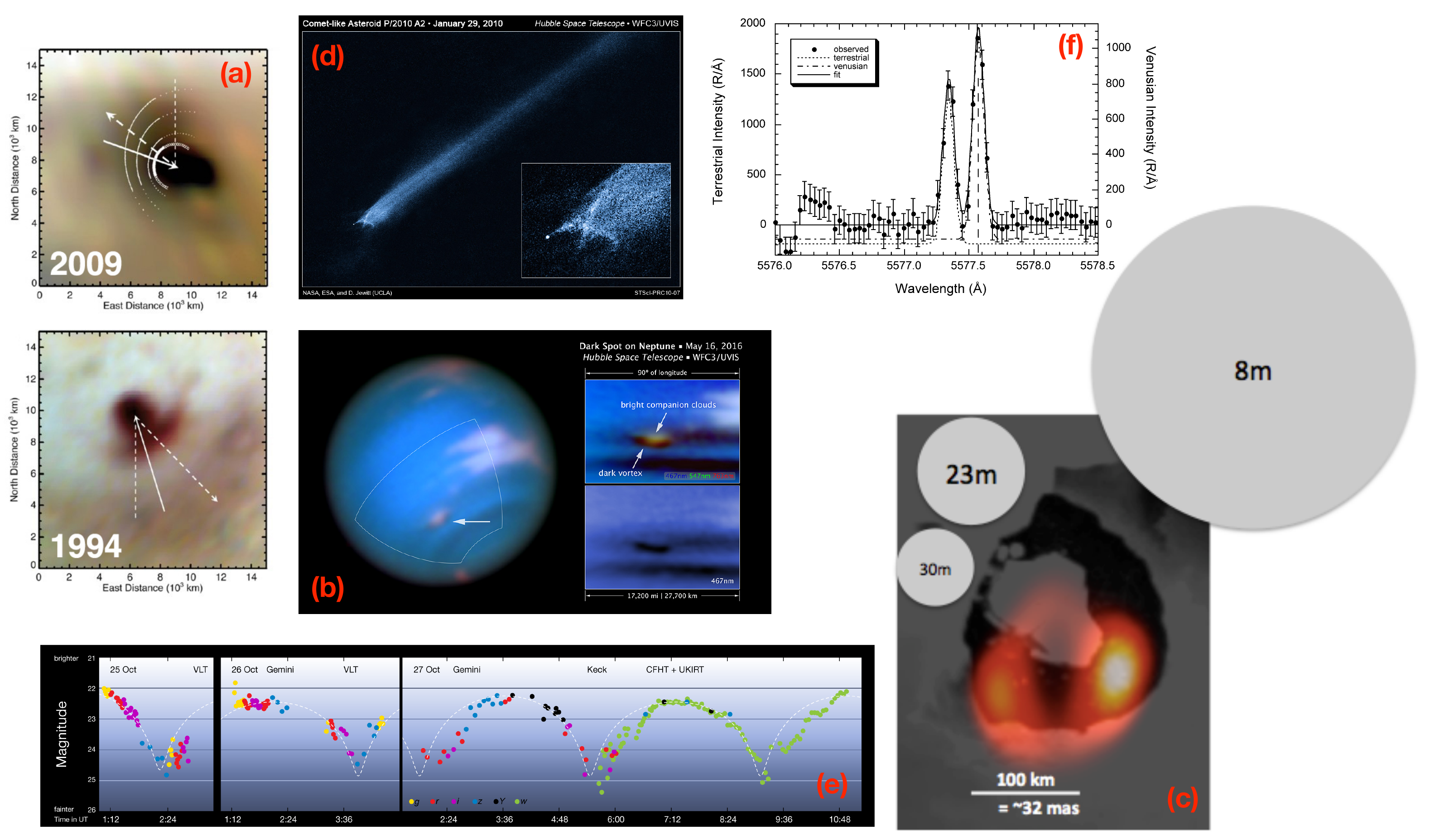}
    \caption{Examples of time-critical solar system phenomena that could be observed through triggered ELT observations. (a) Map projections of Jupiter impact debris acquired after the 2009 impact event (top) and the Shoemaker-Levy 9 impacts in 1994 (bottom) \cite{SanchezLavega2010}.
    (b) Dark vortex on Neptune discovered using HST \cite{Wong2018}. (c) Voyager spacecraft image of a volcanic eruption on Io with emission measured by 23m LBTI overlaid in orange color and smoothed for better visualization.  The resolution disks for that 23m aperture, along with disks for 8 and 30m apertures, are also shown for comparison. (d) HST observation of main belt comet P/2010 A2 revealing an extended dust tail \cite{Jewitt2010}. (e) Light curve of `Oumuamua obtained over three days in October 2017 showing a brightness range of 2.5 magnitudes \cite{eso}. (f) Oxygen green line emission detected on Venus' night side with the Keck 10m telescope shortly after a solar flare and coronal mass ejection \cite{Slanger2006,Gray2014}.}
    \label{fig:montage}
\end{figure}

\textbf{Impacts:} Impacts on giant planets are rare, but have
provided unique insights into impactor populations, the physics of high-velocity atmospheric impacts, and giant planet stratospheric chemistry and circulation on multiple time scales. To date, we have witnessed a total of seven impact events on Jupiter: a fireball recorded by Voyager 1 in 1979 \cite{Cook1981}, the impacts of the fragments of Comet Shoemaker-Levy 9 in 1994 \cite{Harrington2004}, an impact scar first detected by an amateur astronomer in 2009 and then observed with the Hubble Space Telescope \cite{SanchezLavega2010,dePater2010,Hammel2010,Fletcher2010,Orton2011,SanchezLavega2011}, and four subsequent impact flashes detected by amateur astronomers using high-speed imaging techniques \cite{Hueso2010a,Hueso2013,Rogers2010} (Fig.~\ref{fig:montage}a).
Each of these events provided additional constraints on the nature of the impactors, the size distribution of small bodies in the vicinity of Jupiter, and the impact rates. An extrapolation of cratering record on the Galilean satellites suggests that Jupiter should receive $\sim$1 detectable impact ({\it i.e.} from a 10m-class object) every Earth year \cite{Schenk2004}. In contrast, modeling the orbital evolution of known comets and asteroids that encounter Jupiter and extrapolating the result to the 10m-class objects would result in 50 such collisions every year \cite{Levison2000}. The discrepancy between these estimates comes from a lack of data on the size distribution of objects in the 10m class or smaller. As four impacts have been detected since 2009 using modest aperture telescopes, the event rate is likely to be significantly greater than one/year. Accumulating impact records is the only viable way of determining the size distribution of bodies in that class because such bodies are too small to be detected through telescopic observations.\\
The size distribution of small bodies is collisionally controlled; thus, determination of the small body size distribution will shed light on the evolution of the outer solar system.  One scenario that has been relatively successful in explaining the orbital distribution of small bodies in the outer solar system (Trojan asteroids, Kuiper belt), the ``Nice model,'' invokes migration and mean-motion resonance of the outer planets \cite{Morbidelli2005,Tsiganis2005,Gomes2005}. If the generalities of the Nice model are correct, substantial collisional processes must have occurred during the scattering of KBOs, leading to the production of small, 10m-class objects. An analysis of Jovian impacts enabled through ELT observations will add important constraints to these solar system evolution models.\\
\textbf{Outer planet storms:} Superstorm eruptions on Saturn, Uranus, and Neptune are key to understanding the heat transport within hydrogen-dominated atmospheres, particularly the nature of moist convection and its inhibition by molecular weight stratification. Small telescopes equipped with modern digital imaging capabilities rival the capabilities that were available with professional telescopic facilities during the 1965 equinox of Uranus. Even 1m-class telescopes now routinely resolve atmospheric features on Uranus, and these detections can be used to trigger ELT observations to study episodic atmospheric events. If a bright atmospheric feature appears and is confirmed through multiple ground-based observations sufficiently to predict its longitudinal motion, ELT observations are then required in order to a) resolve the detailed vertical and horizontal structure of such a feature, and b) reveal any secondary features such as dark spots that may be formed by a companion storm (Fig.~\ref{fig:montage}b), and elucidate further details of their morphological evolution as driven by atmospheric dynamics.  ELT observations of a new dark spot in the process of its formation would be especially valuable because little is known about the processes that form the dark anticyclones. To date, HST has detected only two dark spots on Uranus, and 5 on Neptune \cite{Hammel2009,Sromovsky2012,Hsu2019,Simon2019}. Also, thermal balance of Uranus and Neptune is poorly understood; in particular, the zero thermal flux of Uranus derived by Voyager \cite{Pearl1990} may have represented a quiescent inter-storm period, while much of the heat flux may be carried by episodic storms.  Thus, ELT observations of transient phenomena on Uranus and Neptune have implications for understanding the radiative-convective balance in their atmospheres.\\
\textbf{Extreme eruptions on Io:} The prodigious volcanic activity of the Jovian satellite Io is a consequence of tidal heating on an eccentric orbit maintained by the 4:2:1 Laplace orbital resonance with Europa and Ganymede \cite{Peale1979}, but the energy dissipation, melt production and transport, and the eruptions themselves are poorly understood \cite{Davies2007}.  Long-term monitoring with 8-10 m telescopes has revealed persistent yet dynamic activity, including occasional “outburst” eruptions that double the radiant flux from Io at 5 $\mu$m \cite{deKleer2016a}.  Observed near inception, these eruptions can be precisely located and eruption temperatures measured; the latter constrains the composition and degree and location of partial melting within Io's interior.  Time-series observations obtain the duration, changes in location, eruption style, and effusion and eruption rates.  These will help us understand the mechanism of Io's volcanism and by inference other bodies with high heat flow such as the terrestrial planets in the early Solar System and volanically active exoplanets.  The greater sensitivity and resolution of TMT and GMT over current 8-10 m telescopes (Io subtends only 1.2" at opposition) will provide more precise spatial and spectral information.  As a demonstration, observations with the Large Binocular Telescope Interferometer \cite{Conrad2015} were able to distinguish emission from two separate active regions within a single lava lake (Fig.~\ref{fig:montage}c).  Since the eruptions vary on timescales of minutes, hours, days, and much longer, interrupt observations with ELTs are critical to observe the site (which often cannot be observed again for several days because  of Io's orbit). Similar investigations can be conducted for  cryovolcanic eruptions or plumes on ocean worlds such as Europa.\\
\textbf{Small body collisions:} The Large Synoptic Survey Telescope (LSST) will detect large numbers of asteroids each night, including objects that exhibit activity due to escape of dust \cite{Ivezic2008,LSSTSciBook2009}. Characterization and followup of active asteroid brightening events was deemed as a high priority science goal by the LSST Solar System Science Collaboration \cite{Schwamb2018}.
There are two proposed mechanisms for the origin of this comet-like activity, which has been observed on several Main Belt asteroids using the Hubble Space Telescope (Fig.~\ref{fig:montage}d). For some objects, the volatile escape reaches a maximum near perihelion and is repeatable, suggesting a comet-like origin for these emissions \cite{Jewitt2012}.  For others, the observed dust emissions are not correlated with heliocentric distance, suggesting that the dust may be a result of collisions between small bodies \cite{Jewitt2011}. There are currently 18 active asteroids (those driven by volatile outgassing are referred to as Main Belt Comets) known \cite{Snodgrass2018}, yet no single explanation for their gas and dust emissions matches all observations. If such objects are discovered through routine LSST observations, they can be used to trigger ELT observations to obtain images and spectra that together can be used to quantify the amount of ice present in these objects, and by extrapolation, in the Main Belt. This has significant astrobiology implications since these objects may represent relics of the population that delivered water to the early Earth.\\
\textbf{Unexpected phenomena}: The discovery of the first interstellar object (ISO) 1I/2017 U1 (`Oumuamua) in October 2017 \cite{Meech2017} has opened a new window on a broader perspective of planetary systems represented by the ejected building blocks and fragments that fortuitously pass through our solar system \cite{Raymond2018}. `Oumuamua was discovered during the course of routine Near Earth Asteroid sky surveys and it moved very quickly through the inner solar system. Within the time it took to mobilize telescope assets to conduct triggered target of opportunity observations it had faded significantly, thus reducing the quality and amount of data that were obtained.\\
LSST is expected to detect more of these objects \cite{Engelhardt2017} and visible and infrared photometry, astrometry, adaptive optics imaging, and spectroscopy are needed to determine the orbit, shape, rotation, presence or absence of a coma, and composition of these objects (Fig.~\ref{fig:montage}e).  The LSST Solar System Science Collaboration has also ranked this topic as a high priority science goal \cite{Schwamb2018}. Not only has `Oumuamua proven to be enigmatic but its investigation was limited by its small size ($\sim$ 100m), faintness ($H_V \approx 22.4$) and a hyperbolic, highly inclined trajectory though the inner Solar System ($e\sim 1.20$, $i=123^{\circ}$).  If `Oumuamua is characteristic of ISOs then effective follow-up will required very large aperture and rapid, triggered response in both geographic hemispheres to characterize these objects.  The exceptional range of photometric variability of `Oumuamua can be explained by an extreme aspect ratio \cite{Meech2017} or binary shape \cite{Gaidos2018} and the shape of ISOs are informative about their formation mechanism and tidal stress history \cite{Raymond2018}.  These objects will not be spatially resolved (100m at 1 au = 0.14 mas) but the shape can be inferred by changes in the rotational light curve as the object moves along its trajectory as aspect and illumination angles change, driving the requirement for precise photometry to distances $\gg 1$au and magnitudes to $V \sim 30$.  Low-resolution optical and infrared spectroscopy of ISOs can distinguish between icy, rocky, and surfaces similar to the classic asteroid types, but current telescopes are not adequate for definitive conclusions \cite{Fitzsimmons2017}.  Fainter surface brightness sensitivity will permit detection or improved upper limits on dusty outgassing \cite{Meech2017,Jewitt2017}.\\

\textbf{Atmospheric response to solar activity}: Observing the effects on terrestrial planet atmospheres of ionizing radiation and charged particles originating from the Sun during solar flares or coronal mass ejections provides an opportunity to study space weather on other planets. This has important implications for understanding the role of stellar activity in driving atmospheric evolution and escape. Large numbers of exoplanets have been and will be discovered around M dwarfs, many of which have high levels of stellar activity. Thus our own solar system serves as an ideal laboratory for advancing our understanding of the role of space weather in atmospheric evolution and by extension the habitability of exoplanets around active stars.\\
Auroral and nightglow oxygen emissions have been detected on the nightside of Venus using 3 and 10m class telescopes \cite{Slanger2006,Gray2014}, but the spatial resolution has been insufficient to characterize their spatial structure. The appearance of the oxygen green line is correlated with solar activity, and it is more closely linked to charged particle precipitation due to coronal mass ejections than enhanced EUV flux from solar flares \cite{Gray2014}. Spatially resolved measurements of Venus' optical and NIR nightglow emissions are critical for understanding the variation of Venus' atmospheric flows with altitude, as the IR O$_2$ emissions are diagnostic of general atmospheric circulation whereas the atomic oxygen green line emission follows the subsolar to antisolar flow.  These observations would be triggered by the detection of major solar eruptions during periods of high solar activity. Observations of Venus' nightglow emissions are optimal when the Earth-Venus relative velocity is maximized, {\it i.e.} the emissions are Doppler shifted out of the terrestrial airglow lines (Fig.~\ref{fig:montage}f).  Observations of Venus made between maximum western elongation and inferior conjunction would be feasible; Venus would be observable during morning twilight with at least half of its night side visible from Earth. ELTs would offer unprecedented spatial resolution of the nightglow emissions on Venus, providing critical information on the atmospheric response to stellar activity.\\

{\bf The science questions addressed by the aforementioned observations are diverse but all can be linked to the fundamental questions concerning the formation and evolution of our solar system. Our solar system remains an important laboratory for understanding physical processes such as migration, impacts, heat balance, and star-planet connections as we begin to develop a comprehensive understanding of the workings of planetary systems overall.}

\pagebreak

\bibliographystyle{abbrv}
\bibliography{refs.bib}

\end{document}